\documentclass[conference]{IEEEtran}

\usepackage[letterpaper, left=1in, right=1in, bottom=1in, top=0.75in]{geometry}
\usepackage{cite}
\usepackage[font=scriptsize]{caption}
\usepackage{subcaption}
\usepackage{algorithm}
\usepackage{algpseudocode}
\makeatletter
\def\BState{\State\hskip-\ALG@thistlm}
\makeatother

\usepackage{mathtools}
\DeclarePairedDelimiter{\ceil}{\lceil}{\rceil}
\usepackage{graphicx}
\usepackage{amsmath}
\usepackage{hyperref}
\usepackage{amssymb}
\usepackage{amsthm}

\newtheorem{prop}{\hspace{0.5cm}Proposition}
\newtheorem*{proof*}{\hspace{0.8cm}{\it{Proof}}}

\ifCLASSINFOpdf

\else

\fi
\begin{document}
\title{ Power Efficient Trajectory Optimization for the Cellular-Connected Aerial Vehicles }
\author{    \IEEEauthorblockN{Behzad Khamidehi and Elvino S. Sousa}
	\IEEEauthorblockA{
		Department of Electrical and Computer Engineering, University of Toronto, Canada}
	Emails: b.khamidehi@mail.utoronto.ca and es.sousa@utoronto.ca}
\maketitle

\begin{abstract}
Aerial vehicles have recently attracted significant attention in a variety of commercial and civilian applications due to their high mobility, flexible deployment and cost-effectiveness.
To leverage these promising features, the aerial users have to satisfy two critical requirements: First, they have to maintain a reliable communication link to the ground base stations (GBSs) throughout their flights, to support command and control data flows. Second, the aerial vehicles have to minimize their propulsion power consumption to remain functional until the end of their mission. In this paper, we study the trajectory optimization problem for an aerial user flying over an area including a set of GBSs. The objective of this problem is to find the trajectory of the aerial user so that the total propulsion-related power consumption of the aerial user is minimized while a cellular-connectivity constraint is satisfied. This problem is a non-convex mixed integer non-linear problem and hence, it is challenging to find the solution. To deal with, first, the problem is relaxed and reformulated to a more mathematically tractable form. Then, using successive convex approximation (SCA) technique, an iterative algorithm is proposed to convert the problem into a sequence of convex problems which can be solved efficiently.

\end{abstract}

%
\IEEEpeerreviewmaketitle

\section{Introduction}
\IEEEPARstart{U}{nmanned} aerial vehicles (UAVs) have shown remarkable potential to improve the performance of a variety of applications in the cellular networks. This is caused by high mobility, flexible deployment, ubiquitous accessibility, robust navigation, and cost effectiveness of the UAVs \cite{R_Zhang_Survey}. To integrate the UAVs into the cellular networks, the following two categories are considered: UAV-assisted cellular communication and cellular-enabled UAV communication \cite{Perspective_RZhang}. While in the former one, the UAVs operate as communication platforms to support the terrestrial users (e.g., aerial base stations (BSs) or relays), in the latter one, the UAVs act as users which are supported by the ground base stations (GBSs). Aerial surveillance, search and rescue, package delivery, photography, and traffic monitoring are some of the most important applications of UAVs operating as aerial users \cite{Perspective_RZhang,Survey_Implementation}.

It is worth mentioning that although the UAVs bring considerable benefits to the cellular networks, the proper design of the network is highly challenging. This difficulty stems mainly from the following two reasons: First, it is crucial for the UAVs to maintain a reliable communication link to the GBSs to control the command and data flows between the UAVs and the cellular network, which affects the safety of the system. Second, since the battery capacity of the UAVs is limited, it is vital to minimize the power consumption of the UAVs to keep them operational until the end of their missions \cite{Energy_tradeoff}. Several studies have been conducted recently to address these issues \cite{Perspective_RZhang,MEC_RZhang,Guevenc,Noma,Saad}. In \cite{Perspective_RZhang}, the authors formulated a UAV trajectory optimization problem to minimize the flight time between a given pair of initial and destination. This problem is subject to a minimum received signal to noise ratio (SNR) constraint to maintain the cellular-connectivity. In \cite{MEC_RZhang}, a computation offloading problem for a cellular-connected UAV is studied. The goal of the problem is to minimize the UAV's mission completion time by optimizing its trajectory and computation offloading scheduling. In \cite{Guevenc}, the authors proposed a dynamic programming approach to optimize the trajectory of an aerial vehicle. The objective is to minimize the time to fly from initial to the destination, ensuring that the disconnection duration constraint for the cellular connectivity is satisfied. It is worth mentioning that although these studies focused on minimizing the UAV's mission completion time, they did not consider the propulsion-related power consumption of the UAV in their problem formulation. This is essential for the UAVs, since not only their power consumption depends on the length of their paths, but also it is a function of the UAV's speed and acceleration. As a result, the UAV's power consumption has to be efficiently minimized so that the UAV remains functional until the end of its mission.    

To address the aforementioned issues, this paper studies the trajectory optimization problem for an aerial vehicle with the objective of minimizing the total propulsion-related power consumption of the aerial vehicle, ensuring that a cellular-connectivity constraint is satisfied. This problem is a non-convex mixed integer problem, and in terms of computational complexity, it is categorized into the class of NP-hard problems. To tackle this problem, we first relax and reformulate the problem into a more mathematically tractable form. Then, based on the successive convex approximation (SCA) technique, we develop an iterative algorithm to convert the original challenging problem into a sequence of convex optimization problems which can be solved efficiently. Simulation results show that the developed algorithm performs well and converges in a few iterations.     

The rest of the paper is organized as follows: Section II presents the system model and formulates the optimization problem. The feasibility of the problem is also discussed in this section. The problem reformulation and the proposed algorithm are presented in section III. Section IV presents the simulation results, and finally, section V concludes the paper.   

\section{System Model}
\begin{figure}[t]
	\centering
	\includegraphics[width=3.3in,keepaspectratio, trim=0cm 0cm 0cm 3cm,]{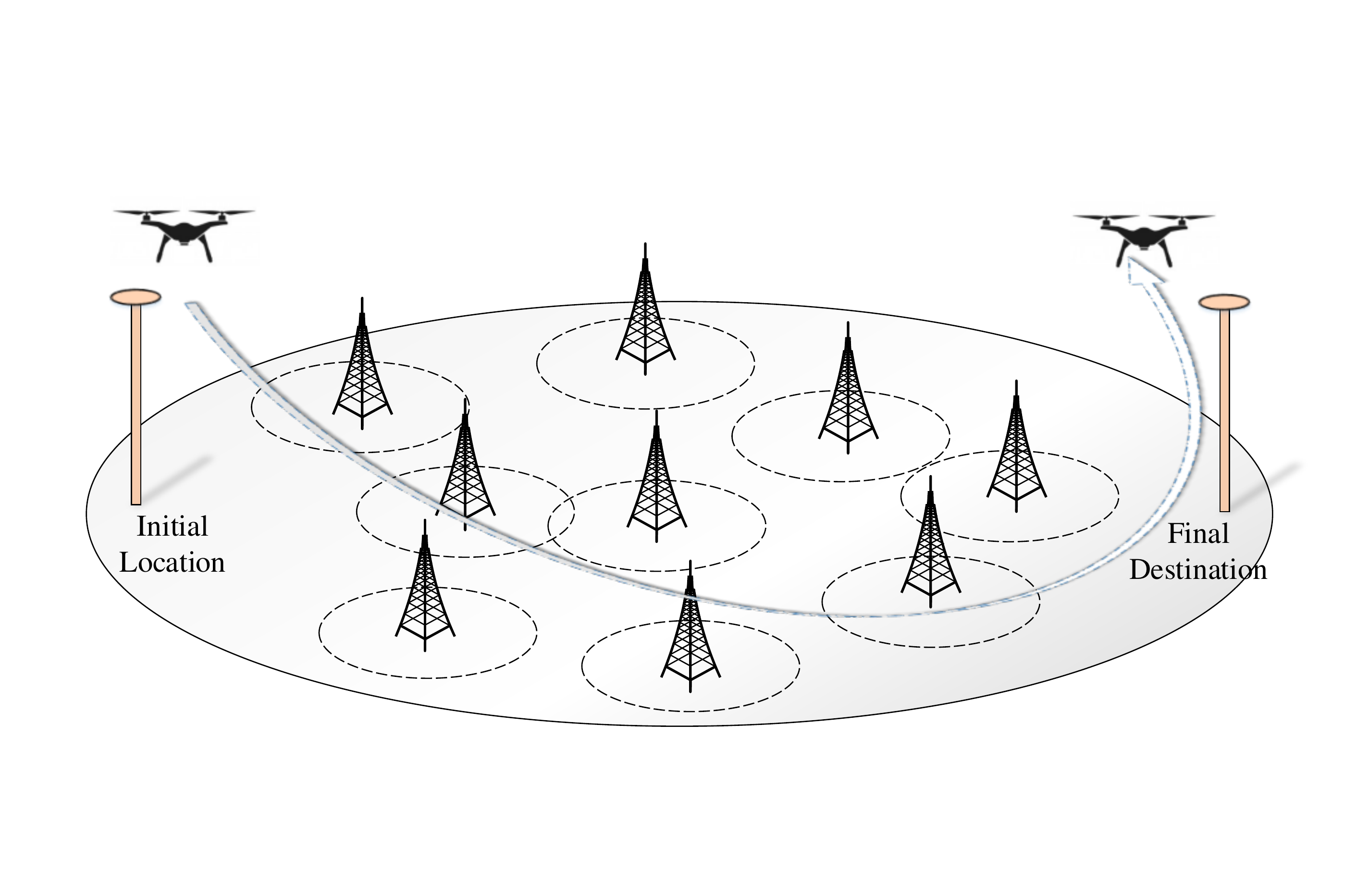}
	\vspace{-1.2cm}
	\caption{An aerial user flying from an initial location to its destination.}
	\label{fig_1}
	\vspace{-1.5em}
\end{figure}

In this paper, as presented in Fig. \ref{fig_1}, we consider an aerial user whose mission is to fly from an initial point to a final destination. The area of the flight consists of $J$ ground base stations. To support the aerial user through its flight, the aerial user is connected to the GBSs. The constraint for the aerial user is, not to loose its cellular connection to the GBSs for more than $T_c$ time units. We denote the 3D position of the aerial user at time $t$ by ${\bf{Q}}(t)=(x(t),y(t),H)$, where $H$ is the altitude of the aerial user. In this work, we assume that the altitude is constant. The maximum flight time of the aerial user is denoted by $T$. To satisfy the cellular-connectivity constraint of the aerial user, it is sufficient to show that it is connected to one of the GBSs every $T_c$ time units. By this, we guarantee that the the maximum time duration that the aerial user is disconnected from the cellular connection is limited to $T_c$. In this regard, we discretize the time interval into $N=\ceil{\frac{T}{T_c}}$ slots as follows: 
\begin{equation*}
\label{time_split}
t_0=0, t_1=T_c, \ldots, t_{N-1}=(N-1)T_c, t_{N}=T.
\end{equation*}   
Using this notation, the position of the aerial user over time $T$ is denoted by $\{{\bf{Q}}[n]\}_{n=1}^{N}$, where ${\bf{Q}}[n] \triangleq {\bf{Q}}(t_n)$. In a similar manner, the speed and the acceleration of the aerial user over time $T$ is represented by $\{{\bf{v}}[n]\}_{n=1}^{N}$ and $\{{\bf{a}}[n]\}_{n=1}^{N}$, respectively. If the location of the $j$-th GBS is denoted by ${\bf{Q}}_j$, the distance between the aerial user and the $j$-th GBS at time $t_n$ will be described as 
\begin{equation}
\label{Distance}
d_j [n]=\rVert {\bf{Q}} [n] - {\bf{Q}}_j \rVert.
\end{equation}
We assume that the communication links between the aerial user and the GBSs are dominated by the line-of-sight (LoS) link \cite{TWC_joint_R_Zhang}. Accordingly, the channel gain between the aerial user and the $j$-th GBS  at time $t_n$ is given by
\begin{equation}
\label{channel}
g_j [n] = \frac{\beta_0}{d^2_j [n]}=\frac{\beta_0}{\rVert {\bf{Q}} [n] - {\bf{Q}}_j \rVert^2},
\end{equation}
where $\beta_0$ is the channel gain at the reference distance $d=1$m. We also define the user association indicator as 
\begin{equation}
\label{user_association}
\alpha_j [n] \in \{0,1\}, \forall j,n,
\end{equation} 
where $\alpha_j [n]=1$ indicates that at time $t_n$, the aerial user is served by GBS $j$. Otherwise, $\alpha_j [n]=0$. Let $P_j [n]$ denote the tranmit power of the $j$-th GBS at time $t_n$. If the aerial user is served by the $j$-th GBS at time $t_n$, the correponding received signal to interference plus noise ratio (SINR) at the aerial user is given by
\begin{equation}
\label{SINR}
\gamma_{j} [n] = \frac{ \frac{h_j [n]}{\rVert {\bf{Q}} [n] - {\bf{Q}}_j \rVert^2}        }{\displaystyle \sum_{j'\neq j, j'=1}^{J} \frac{h_{j'} [n]}{\rVert {\bf{Q}} [n] - {\bf{Q}}_{j'} \rVert^2}  + 1},
\end{equation}
where $h_k [n]=\frac{P_k[n] \beta_0}{\sigma^2}$, $\forall k$, is the referene SNR for the $k$-th GBS, and $\sigma^2$ is the noise power at the receiver. 

\subsection{Problem Formulation}

As discussed earlier, the goal of this work is to obtain the trajectory of the aerial user such that the propulsion-related power consumption of the aerial user is minimized, ensuring that the cellular-connectivity constraint of the aerial user is satisfied. In what follows, the trajectory optimization problem is formulated.

The total propulsion-related power consumption of the aerial user is given by   
\begin{equation}
\label{Power_AU}
P ({\bf{v}}, {\bf{a}} )= \displaystyle  \sum_{n=1}^{N} c_{1} \rVert {\bf{v}} [n] \rVert ^3 +\frac{c_{2}}{\rVert {\bf{v}} [n] \rVert} \left (  1+\frac{\rVert {\bf{a}} [n] \rVert^2 }{g} \right),
\end{equation}
where $c_1$ and $c_2$ are constants depending on the air density, wing area, drag coefficient, etc \cite{TWC_Energy_Efficient_R_Zhang}, and $g$ is the gravitational acceleration. The constraints of the optimization problem are as follows:
\begin{itemize}
	\item To satisfy the cellular-connectivity constraint of the aerial user, the received SINR by the aerial user at each time $t_n$ must be greater than the minimum SINR required for the communication between the aerial user and the GBSs denoted by $\gamma_{\text{min}}$. This can be expressed as
	\begin{equation}
	\label{SINR_constraint}
	\gamma_{j} [n] \geq \gamma_{\text{min}} \alpha_{j} [n], \hspace{0.5cm}\forall n, j.
	\end{equation}   
	It is worth mentioning that at time $t_n$, the cellular-connectivity constraint has to be met only for the link between the aerial user and its serving GBS. Due to this fact, in the right side of  (\ref{SINR_constraint}), $\alpha_{j} [n]$ is multiplied to emphasize that this is valid only for the GBS serving the aerial user.
	\item At each time $t_n$, the aerial user has to be served by one GBS, i.e.,
	\begin{equation}
	\label{association_constraint}
	\displaystyle \sum_{j=1}^{J} \alpha_{j} [n]  =1, \hspace{0.5cm} \forall n.
	\end{equation}
	\item The trajectory of the aerial user is subject to the following equations:
	 \begin{equation}
	{\bf{Q}} [n] = {\bf{Q}} [n-1] + {\bf{v}} [n-1] T_c + \frac{1}{2} {\bf{a}} [n-1] T_c^2, 
	\end{equation}
	\begin{equation}
	{\bf{v}} [n] = {\bf{v}} [n-1] + {\bf{a}}[n-1]  T_c,
	\end{equation}
	\begin{equation}
	{\bf{Q}} [0] ={\bf{Q}}^0, {\bf{Q}} [N+1] ={\bf{Q}}^F, \text{ and }  {\bf{v}} [0] = {\bf{v}}^0, 
	\end{equation}
	where ${\bf{Q}}^0$, ${\bf{Q}}^F$ are the initial and final location of the aerial user, respectively, and ${\bf{v}}^0$  is the initial speed.
	\item If ${\bf{v}}_{\text{max}}$ and ${\bf{a}}_{\text{max}}$ denote the maximum speed and the maximum acceleration of the aerial user, respectively, we have 
	 \begin{equation}
	\rVert {\bf{v}} [n]  \rVert \leq {\bf{v}}_{\text{max}},
	\end{equation}
	\begin{equation}
	\rVert {\bf{a}} [n]  \rVert \leq {\bf{a}}_{\text{max}}.
	\end{equation}
\end{itemize}

If ${\bf{Q}}= \{{\bf{Q}}_j [n], \forall j,n\}$, ${\bf{v}}=\{{\bf{v}}_j [n], \forall j,n\}$, ${\bf{a}}=\{{\bf{a}}_j [n], \forall j,n\}$, and $\alpha=\{\alpha_j [n], \forall j,n\}$, according to the aforementioned constraints, the optimization problem can be formulated as 
\begin{subequations}
	\begin{align}
	\label{Problem_original}
	\min_{{\bf{Q}}, {\bf{v}}, {\bf{a}}, {\bf{\alpha}}}  & P ({\bf{v}}, {\bf{a}} ) \\ 
	\text{s.t. }  & \begin{matrix} \label{SINR_constraint_1}  \gamma_{j} [n] \geq \gamma_{\text{min}} \alpha_{j} [n], \hspace{0.5cm}\forall n, j, \end{matrix} \\
	& \begin{matrix}\label{association_constraint_1} \displaystyle \sum_{j=1}^{J} \alpha_{j} [n]  =1, \hspace{0.5cm}\forall n, \end{matrix} \\
	& \begin{matrix} \label{ Q_cons} {\bf{Q}} [n] = {\bf{Q}} [n-1] + {\bf{v}} [n-1] T_c + \\ \frac{1}{2} {\bf{a}} [n-1] T_c^2, \hspace{0.5cm}\forall n,\end{matrix}\\
	& \begin{matrix} \label{v_cons }
		{\bf{v}} [n] = {\bf{v}} [n-1] + {\bf{a}}[n-1]  T_c, \hspace{0.5cm}\forall n,
	\end{matrix}\\
	& \begin{matrix} \label{init_final}
	{\bf{Q}} [0] ={\bf{Q}}^0, {\bf{Q}} [N+1] ={\bf{Q}}^F, {\bf{v}} [0] = {\bf{v}}^0,
	\end{matrix}\\
	& \begin{matrix} \label{v_limits}
	\rVert {\bf{v}} [n]  \rVert \leq {\bf{v}}_{\text{max}}, \hspace{0.5cm} \forall n,
	\end{matrix}\\
	& \begin{matrix} \label{a_limits }
	\rVert {\bf{a}} [n]  \rVert \leq {\bf{a}}_{\text{max}}, \hspace{0.5cm} \forall n,
	\end{matrix}\\
	& \begin{matrix} \label{binary_alpha}
	\alpha_j [n] \in \{0,1\}, \hspace{0.5cm} \forall j,n,
	\end{matrix}
	\end{align}
\end{subequations}

\subsection{Feasibility of the Problem}
The feasibility of problem (\ref{Problem_original}) depends on the topology of the network and the location of the GBSs. In this section, we briefly discuss the feasibility of (\ref{Problem_original}). To this goal, first, we introduce some auxiliary variables as follows: $I_j [n]\triangleq \displaystyle \sum_{j'\neq j, j'=1}^{J} \frac{h_{j'} [n]}{\rVert {\bf{Q}} [n] - {\bf{Q}}_{j'} \rVert^2}$, 
$r_j^2 [n] \triangleq  \left( \frac{P_j [n] \beta_0}{\gamma_{\text{min}} (I_j [n] + \sigma^2)} - H^2 \right)$, $\forall n,j$, and $d_0 \triangleq {\bf{v}}_{\text{max}} T_c$. Moreover, we define association vector as ${\bf{K}} \triangleq [k_1, k_2, \ldots, k_N]$, indicating that the aerial user is served by GBS $k_1$ at time $t_1$, then served by GBS $k_2$ at time $t_2$, and so on. Using these notations, the problem (\ref{Problem_original}) has a feasible solution if there exists an association vector satisfying the following constraints:
\begin{equation*}
\rVert {\bf{Q}}_{k_1} -{\bf{Q}}^0 \rVert \leq r_{k_1}[1] + d_0,
\end{equation*} 
\begin{equation*}
\rVert {\bf{Q}}_{k_n} -{\bf{Q}}_{k_{n-1}} \rVert \leq r_{k_n}[n] + d_0 + r_{k_{n-1}}[n-1], 2 \leq n \leq N,
\end{equation*} 
\begin{equation*}
\rVert {\bf{Q}}_{k_N} -{\bf{Q}}^F \rVert \leq r_{k_N}[N] + d_0.
\end{equation*} 

In this paper, we assume that the topology of the network is such that all these constraints are satisfied by some association vectors, and hence, the problem has a feasible solution.

\section{Proposed Solution}
The optimization problem (\ref{Problem_original}) is a non-convex mixed integer problem due to the following reasons: First, the variables $\{\alpha_j [n]\}_{n=1}^{N}$ that are involved in constraints (\ref{SINR_constraint_1}) and (\ref{association_constraint_1}) are binary. Second, the objective function and constraint (\ref{SINR_constraint_1}) are not convex with respect to the optimization variables (even with a binary $\alpha$). As a result, it is challenging to solve this optimization problem. To make this problem tractable, first we express the binary constraint of (\ref{binary_alpha}) as the intersection of the following continuous constraints:
\begin{subequations}
\begin{align}
& \label{relaxed} 0 \leq \alpha_j [n] \leq 1, \forall j,n, \\
& \label{relaxed2} \displaystyle \sum_{j=1}^{J} \sum_{n=1}^{N} \left( \alpha_j [n] - \alpha_j^2 [n] \right) \leq 0.
\end{align}
\end{subequations}
Using this, the binary variables $\alpha_j [n]$ are relaxed to be continuous in the interval of $[0,1]$. However, we have to consider the fact that any non-integer solution for $\alpha_j [n]$ is not feasible for the original optimization problem. In other words, our goal is to find integer solution for $\alpha_j [n]$, even after the relaxation. To achieve this, we add a cost function to the objective function to penalize it if the values of $\alpha_j [n]$ are not integer. The new objective is expressed as
\begin{equation}
\label{New_obj}
\mathcal{L} ({\bf{Q}}, {\bf{v}},{\bf{a}}, \alpha) = P ({\bf{v}}, {\bf{a}} ) + \lambda \hspace{-0.1cm} \left( \displaystyle \sum_{n=1}^{N} \sum_{j=1}^{J} \left( \alpha_j [n] - \alpha_j^2 [n] \right) \hspace{-0.15cm} \right)\hspace{-0.05cm},
%
%
\end{equation}
where $\lambda$ is the penalty factor. The new optimization problem is formulated as
\begin{subequations}
	\begin{align}
	\label{Problem_new_obj}
	\min_{{\bf{Q}}, {\bf{v}}, {\bf{a}}, {\bf{\alpha}}}  & \mathcal{L} ({\bf{Q}}, {\bf{v}},{\bf{a}}, \alpha)  \\
	\text{s.t. }  & (\ref{SINR_constraint_1}), (\ref{association_constraint_1}), (\ref{ Q_cons}),(\ref{v_cons }), (\ref{init_final}), (\ref{v_limits}), (\ref{a_limits }), \nonumber\\
	&  (\ref{relaxed}). \nonumber
	\end{align}
\end{subequations}
Based on what has been presented in \cite{Behzad_CommL}, it can be shown that for a sufficiently large value of $\lambda$, the optimization problem of (\ref{Problem_original}) is equivalent to (\ref{Problem_new_obj}). 
 Hence, instead of (\ref{Problem_original}), we can solve the problem of (\ref{Problem_new_obj}). It is worth mentioning that although (\ref{Problem_new_obj}) is relaxed and does not include any integer variable, it is still non-convex due to its objective function and also constraint (\ref{SINR_constraint_1}). In general, there is no standard method to solve this problem efficiently. In what follows, using successive convex approximation method, we propose an iterative algorithm to solve the problem of (\ref{Problem_new_obj}). Let $t$ denote the iteration number. In our algorithm, at the $t$-th iteration, the non-convex problem of (\ref{Problem_new_obj}) is approximated with a convex optimization problem. Then, the solution of this convex problem is obtained and it is used as the input in the next iteration to convexify the problem. 
 
 The objective function of (\ref{Problem_new_obj}) is not convex with respect to ${\bf{v}}[n]$. Moreover, the constraint of (\ref{SINR_constraint_1}) is not convex with respect to the trajectory variables ${\bf{Q}}[n]$. To tackle these issues, we introduce the slack variables $\theta [n] \triangleq \rVert {\bf{v}} [n] \rVert$, $\forall n$, and $\rho_j [n] \triangleq \rVert {\bf{Q}}[n] - {\bf{Q}}_j \rVert ^2$, $\forall j,n$. Using these variables, the optimization problem of (\ref{Problem_new_obj}) can be expressed as 
\begin{subequations}
	\begin{align}
	\label{Problem_app_1}
	\min_{{\bf{Q}}, {\bf{v}}, {\bf{a}}, {\bf{\alpha}},\theta, \rho}  & \mathcal{L} ({\bf{Q}}, {\bf{v}},{\bf{a}}, \alpha, \theta)  \\
	\text{s.t. }  & \label{SINR_3} f_{j} (\rho [n])  \geq   \frac{\gamma_{\text{min}}}{h_j [n]}  \alpha_{j} [n] \rho_{j} [n], \\
	& \label{v_theta} \rVert {\bf{v}}[n]  \rVert^2 \geq {\bf{\theta}}^2 [n], \forall n, \\
	& \label{rho_Q} \rho_j [n] = \rVert {\bf{Q}}[n] - {\bf{Q}}_j \rVert ^2, \forall j,n, \\
	& (\ref{association_constraint_1}), (\ref{ Q_cons}), (\ref{v_cons }), (\ref{init_final}),(\ref{v_limits}), (\ref{a_limits }), (\ref{relaxed}), \nonumber
	\end{align}
\end{subequations}
 where
 \begin{equation*}
 \label{Objective_2}
 \mathcal{L} ({\bf{Q}}, {\bf{v}},{\bf{a}}, \alpha, \theta) \hspace{-0.1cm}= \hspace{-0.1cm} P({\bf{v}},{\bf{a}},\theta) 
 + \lambda  \displaystyle \sum_{n=1}^{N} \hspace{-0.05cm} \sum_{j=1}^{J} \left( \alpha_j [n] \hspace{-0.05cm}-\hspace{-0.05cm} \alpha_j^2 [n] \right), 
 \end{equation*}
  \begin{equation*}
  P({\bf{v}},{\bf{a}},\theta)=\displaystyle  \sum_{n=1}^{N} c_{1} \rVert {\bf{v}} [n] \rVert ^3 +\frac{c_{2}}{\theta [n]} \left (  1+\frac{\rVert {\bf{a}} [n] \rVert^2 }{g} \right),
  \end{equation*}
 and 
 \begin{equation*}
 \label{SINR_1}
 f_{j} (\rho [n])= \left(  \displaystyle \sum_{ j'=1, j'\neq j}^{J} \frac{h_{j'} [n]}{\rho_{j'} [n] } +1  \right) ^{\hspace{-0.15cm} -1}.
 \end{equation*}
 The objective function $\mathcal{L} ({\bf{Q}}, {\bf{v}},{\bf{a}}, \alpha, \theta)$ is convex with respect to ${\bf{v}}$ and $\theta$. However, it is still non-convex with respect to $\alpha$. Let ${\bf{Q}}^{t-1}, {\bf{v}}^{t-1},{\bf{a}}^{t-1}, \alpha^{t-1}, \theta^{t-1}$, and $\rho^{t-1}$ denote the solutions of the problem at iteration $(t-1)$. We know that any convex function is understimated by its first order Taylor approximation \cite{convexOpt}. Therefore, to make the objective function of the problem of iteration $t$ convex, we employ the first order Taylor approximation as
\begin{multline}
\label{Objective_3}
\tilde{\mathcal{L}} ({\bf{Q}}, {\bf{v}},{\bf{a}}, \alpha, \theta) = P({\bf{v}},{\bf{a}},\theta) + \lambda  \displaystyle \sum_{n=1}^{N} \sum_{j=1}^{J} \alpha_j [n] \\
- \lambda \Big( \displaystyle \sum_{n=1}^{N} \sum_{j=1}^{J} (\alpha^{t-1}_j [n])^2 + 2 \alpha^{t-1}_j [n] (\alpha_{j}[n]- \alpha^{t-1}_j [n] ) \Big). 
\end{multline} 
 Moreover, the new imposed constraints (\ref{v_theta}) and (\ref{rho_Q}) are not convex. To resolve this issue, using the Taylor approximation, we modify them as  
  \begin{equation}
 \label{v_theta_taylor}
 \rVert {\bf{v}}^{t-1}[n] \rVert^2 + 2 {\bf{v}}^{t-1} [n] ^T \big({\bf{v}} [n] - {\bf{v}}^{t-1} [n] \big) \geq {\bf{\theta}}^2 [n],
 \end{equation}
 and
  \begin{multline}
  \label{rho_Q_taylor}
 {\bf{\rho}}_{j}[n] = \rVert {\bf{Q}}^{t-1} [n] - {\bf{Q}}_j \rVert ^2  \\+ 2 \big({\bf{Q}}^{t-1} [n]- {\bf{Q}}_j \big)^T \big({\bf{Q}} [n]-{\bf{Q}}^{t-1} [n]\big).
 \end{multline}
 To address the convexity of constraint (\ref{SINR_3}), first, we have to investigate concavity of function $f_j (\rho [n])$. The following proposition shows that this function is a concave function.
 \begin{prop}
 	\normalfont The function $f_j (\rho [n])$ is a concave function with respect to variables $\{ \rho_k [n]\}_{\substack{k=1, k \neq j}}^{J}$.
 \end{prop}
 \begin{proof*}
 \normalfont To show that a function is concave, it is sufficient to show that its Hessian matrix is negative-semi-definite. If we define ${\bf{w}}=\left[\frac{h_1 [n]}{\rho_1^2 [n]}, \frac{h_2 [n]}{\rho_2^2 [n]}, \ldots, \frac{h_J [n]}{\rho_J^2 [n]}\right]^T $
 and
 \begin{equation*}
 \Psi \hspace{-0.05cm}= \hspace{-0.05cm} \left(\displaystyle \sum_{ \substack{k=1 \\ k \neq j}}^{J} \frac{h_k [n]}{\rho_k [n]} +\hspace{-0.05cm} 1 \hspace{-0.1cm}   \right)^{ \hspace{-0.05cm}-1} \hspace{-0.3cm} \text{diag} \hspace{-0.05cm} \left \{ \hspace{-0.05cm}\frac{h_1 [n]}{\rho_1^3 [n]}, \frac{h_2 [n]}{\rho_2^3 [n]}, \ldots, \frac{h_J [n]}{\rho_J^3 [n]} \right \},
 \end{equation*}
 we can write the Hessian of function $f_j (\rho [n])$ as
 \begin{multline*}
 \nabla^2 f_j = 2 \hspace{-0.1cm}\left(\displaystyle \sum_{\substack{k=1 \\ k \neq j}}^{J} \frac{h_k [n]}{\rho_k [n]} +1 \hspace{-0.1cm}   \right)^{\hspace{-0.15cm} -3} \times \\
 \hspace{-0.15cm}\left( \hspace{-0.1cm} {\bf{w}}{\bf{w}}^T \hspace{-0.05cm} - \hspace{-0.05cm} \left(\displaystyle \sum_{\substack{k=1 \\ k \neq j}}^{J} \frac{h_k [n]}{\rho_k [n]} +1 \hspace{-0.1cm}   \right) \hspace{-0.1cm} \Psi \hspace{-0.1cm} \right).
 \end{multline*}
 To show that the Hessian is negative-semi-definite, it is sufficient to show ${\bf{y}}^T \nabla^2 f_j {\bf{y}} \leq 0$, for $\forall {\bf{y}} \in R^{J-1}$. We have
 \begin{multline}
 \label{Proof}
 {\bf{y}}^T \nabla^2 f_j {\bf{y}} = 2 \left(\displaystyle \sum_{\substack{k=1 \\ k \neq j}}^{J} \frac{h_k [n]}{\rho_k [n]} +1    \right)^{ \hspace{-0.1cm}-3} \hspace{-0.15cm}{ \left(\displaystyle \sum_{\substack{j'=1 \\ j' \neq j}}^{J} y_{j'} \frac{h_{j'} [n]}{\rho_{j'}^2 [n]}\right )^2   } \\
 - \left( \displaystyle \sum_{\substack{k=1 \\ k \neq j}}^{J} \frac{y_k^2 h_k [n]}{\rho_k^3 [n]}  \right) \left(\displaystyle \sum_{\substack{j'=1 \\ j' \neq j}}^{J} \frac{h_{j'} [n]}{\rho_{j'} [n]} +1    \right) 
 \end{multline}
 If we define vectors $\mu$ and $\nu$ whose components are $\mu_k=\sqrt{\frac{h_k [n]}{\rho_k [n]}}$ and $\nu_k=\frac{\sqrt{h_k [n]}}{\rho_k [n] \sqrt{\rho_k [n]}}y_k$, (\ref{Proof}) can be written as 
 \begin{multline}
 {\bf{y}}^T \nabla^2 f_j {\bf{y}} = 2 \left(\displaystyle \sum_{\substack{k=1 \\ k \neq j}}^{J} \frac{h_k [n]}{\rho_k [n]} +1    \right)^{-3}  \times \\
 \big( (\mu^T \nu)^2 -(\mu^T \mu +1)(\nu ^T \nu) \big).
 \end{multline} 
 According to the Cauchy-Schwartz inequality $(\mu^T \nu)^2 \leq (\mu^T \mu )(\nu^T \nu)$. As a result, ${\bf{y}}^T \nabla^2 f_j {\bf{y}} \leq 0$, and hence, the function $f_j$ is concave. \hfill $\blacksquare$
\end{proof*}
Although function $f_j (\rho)$ is concave, the constraint of (\ref{SINR_3}) is non-convex due to the product term of $\alpha_{j}[n] \rho_{j} [n]$ in the right side. This product can be expressed as
 \begin{equation}
 \alpha_{j}[n] \rho_{j} [n]= \frac{  \left( \alpha_{j}[n] + \rho_{j}[n] \right)^2 - \left(\alpha_{j}^2 [n] + \rho_{j}^2 [n]    \right)  }{2}. 
 \end{equation}
 Since $\left( \alpha_{j}[n] + \rho_{j}[n] \right)^2$ is a convex function, it is always lower bounded by its first order Taylor approximation as
 \begin{multline}
 \alpha_{j}^2 [n] + \rho_{j}^2 [n]   \geq \left(\alpha_{j}^{(t-1)} [n]\right)^2 + \left(\rho_{j}^{(t-1)} [n] \right)^2  \\
 +2 \alpha_{j}^{(t-1)} [n]  \left( \alpha_{j} [n] -\alpha_{j}^{(t-1)} [n] \right) \\
 +2 \rho_{j}^{(t-1)} [n] \left( \rho_{j} [n] -\rho_{j}^{(t-1)} [n] \right) \\
 \triangleq g_{j}^{(t-1)} \left( \rho_{j} [n],\alpha_{j} [n] \right).
 \end{multline}
 As a result, to meet (\ref{SINR_3}), it is sufficient to satisfy the following constraint  
 \begin{equation}
 \label{SINR_final}
 f_j (\rho) \hspace{-0.05cm} \geq \hspace{-0.05cm} \frac{\gamma_{\text{min}}}{2 h_k[n]} \hspace{-0.05cm} \left( \hspace{-0.05cm} \left( \alpha_{j}[n] \hspace{-0.05cm}+\hspace{-0.05cm} \rho_{j}[n] \right)^2 \hspace{-0.05cm}- \hspace{-0.05cm}  g_{j}^{(t-1)} \left( \rho_{j} [n],\alpha_{j} [n] \right) \hspace{-0.06cm} \right)\hspace{-0.1cm}.
 \end{equation}
 Therefore, at iteration $t$, instead of non-convex problem of (\ref{Problem_original}), we can solve the following convex problem
\begin{subequations}
	\begin{align}
	\label{Problem_final}
	\min_{{\bf{Q}}, {\bf{v}}, {\bf{a}}, {\bf{\alpha}},\theta, \rho}  & \tilde{\mathcal{L}} ({\bf{Q}}, {\bf{v}},{\bf{a}}, \alpha, \theta)  \\
	\text{s.t. }  & (\ref{SINR_final}), (\ref{v_theta_taylor}), (\ref{rho_Q_taylor}), \nonumber \\
	& (\ref{association_constraint_1}), (\ref{ Q_cons}), (\ref{v_cons }), (\ref{init_final}), (\ref{v_limits}), (\ref{a_limits }), (\ref{relaxed}). \nonumber
	\end{align}
\end{subequations}

Algorithm \ref{ALg} shows our approach for solving the optimization problem of (\ref{Problem_final}).

	\begin{algorithm}[t]
	\caption{Power-efficient trajectory optimization.}
	\label{ALg}
	\begin{algorithmic}[1]
		\State Initialize $t=0$ and feasible ${\bf{Q}}^0$, ${\bf{v}}^0$, ${\bf{a}}^0$, ${\bf{\alpha}}^0$, ${\bf{\theta}}^0$, and ${\bf{\rho}}^0$.
		\Repeat 
		\State $t=t+1$
		\State Update the objective function and constraints according to \eqref{Objective_3} and (\ref{SINR_final}), (\ref{v_theta_taylor}), and (\ref{rho_Q_taylor}), respectively.
		\State Solve the optimization problem of \eqref{Problem_final} to find the optimal solutions ${\bf{Q}}^t$, ${\bf{v}}^t$, ${\bf{a}}^t$, ${\bf{\alpha}}^t$, ${\bf{\theta}}^t$, and ${\bf{\rho}}^t$.
		\Until The fractional increase of the objective function in \eqref{Problem_final} is less than $\epsilon$.
	\end{algorithmic}
\end{algorithm}

\section{Simulation Results}

In this section, simulation results are presented to show the performance of the proposed algorithm. We consider two maps with different number of GBSs. In the first map, we have $5$ GBSs, while the second environment consists of $8$ GBSs. We assume that all GBSs transmit with the same power, i.e, $P_j [n]=P_0$, $\forall j,n$. The reference SNR $h_k [n]=80$dB, $\forall k,n$. The maximum speed of the aerial vehicle is $15$$\frac{\text{m}}{\text{s}}$ for map $1$, and $12$$\frac{\text{m}}{\text{s}}$ for map $2$. The initial speed of the UAV is ${\bf{v}}^0=(2,2)$$\frac{\text{m}}{\text{s}}$. The maximum acceleration of the aerial vehicle is assumed to be $5$$\frac{m}{\text{s}^2}$. We assume that the aerial vehicle parameters are $c_1=0.002$ and $c_2=80$. The altitude of the UAV is $H=50$m. The gravitational acceleration is $g=10$$\frac{m}{\text{s}^2}$. The flight time of the aerial vehicle is limited to $T=50$s, and the maximum time duration that the aerial vehicle can be disconnected from the cellular connection is $T_c=5$s. The penalty factor in our simulations is set to $\lambda=10^5$.

\begin{figure}
	\centering
	\includegraphics[width=2.7in,keepaspectratio]{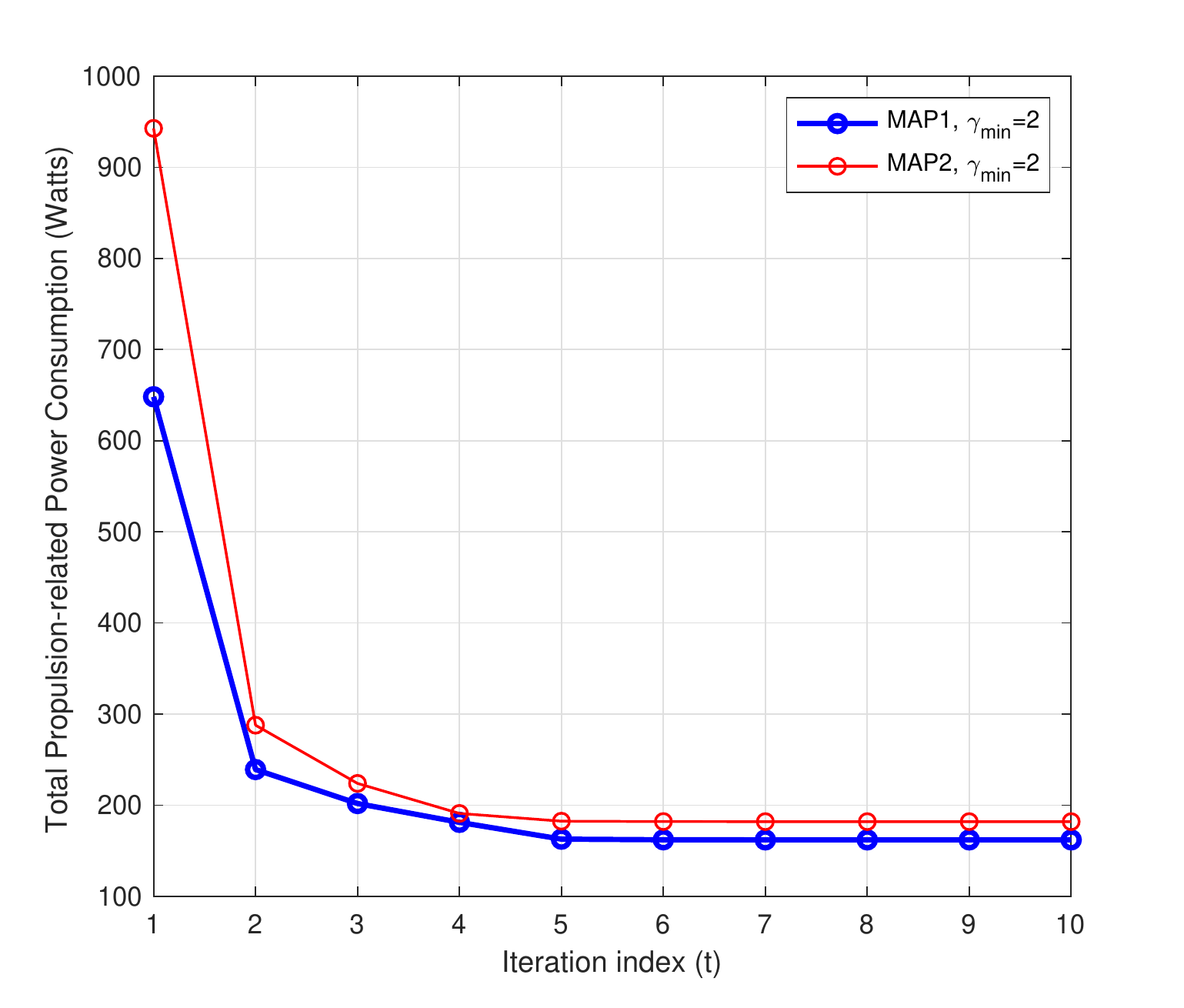}
	\vspace{-0.3em}
	\caption{Convergence of the proposed algorithm when $\gamma_{\text{min}}=2$. }
	\label{convergence}
	\vspace{-1.5em}
\end{figure}%

Fig. \ref{convergence} shows the convergence of the total propulsion-related power consumption of the aerial user for both maps. As can be seen, our proposed algorithm converges in a few iterations. This proves that the algorithm works well and can be implemented in practice. 

\begin{figure}
	\centering
	\includegraphics[width=2.7in,keepaspectratio]{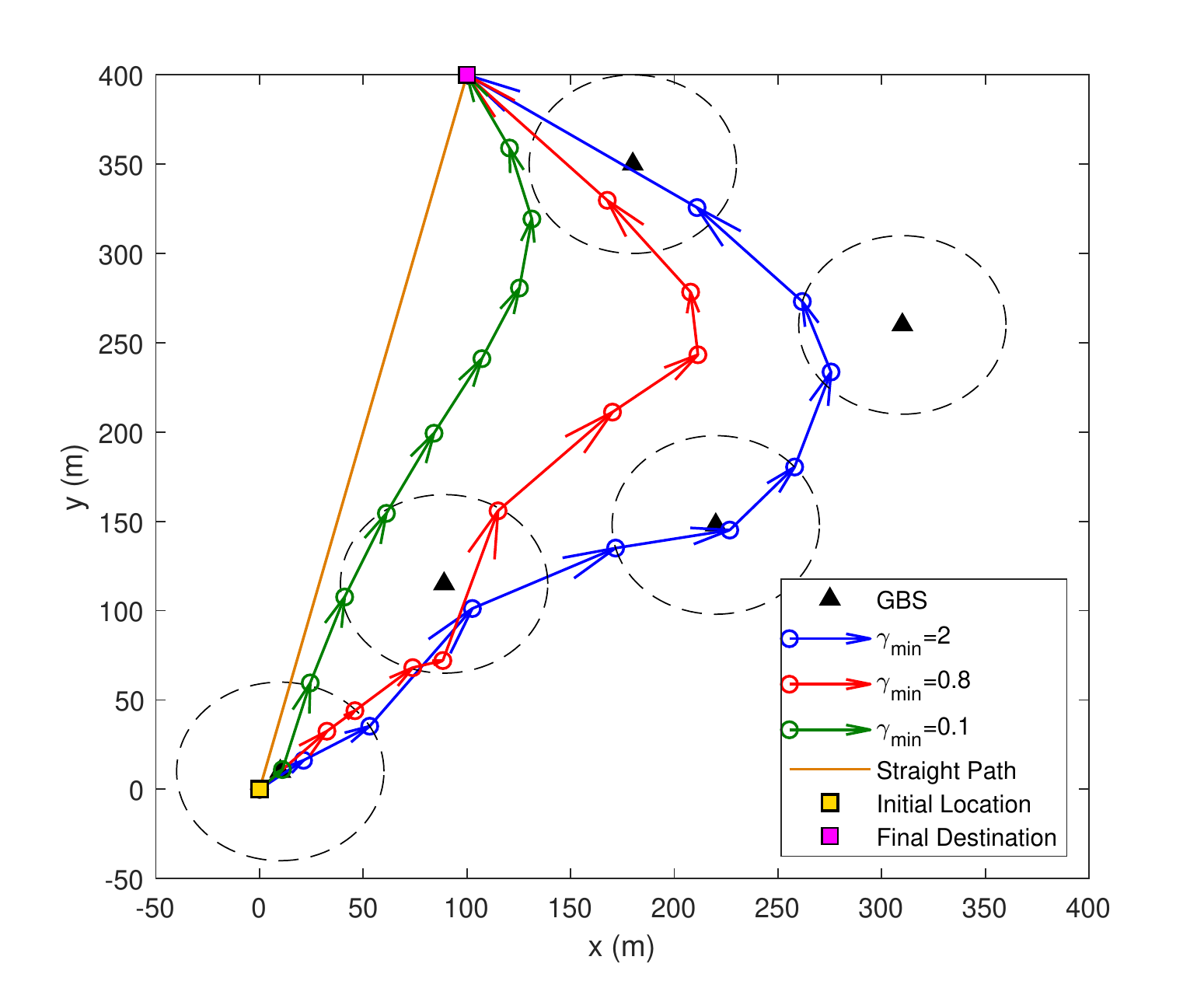}
	\vspace{-0.3em}
	\caption{Trajectory of the aerial vehicle for map 1.}
	\label{traj_1}
	\vspace{-1.5em}
\end{figure}%

\begin{figure}
	\centering
	\includegraphics[width=2.7in,keepaspectratio]{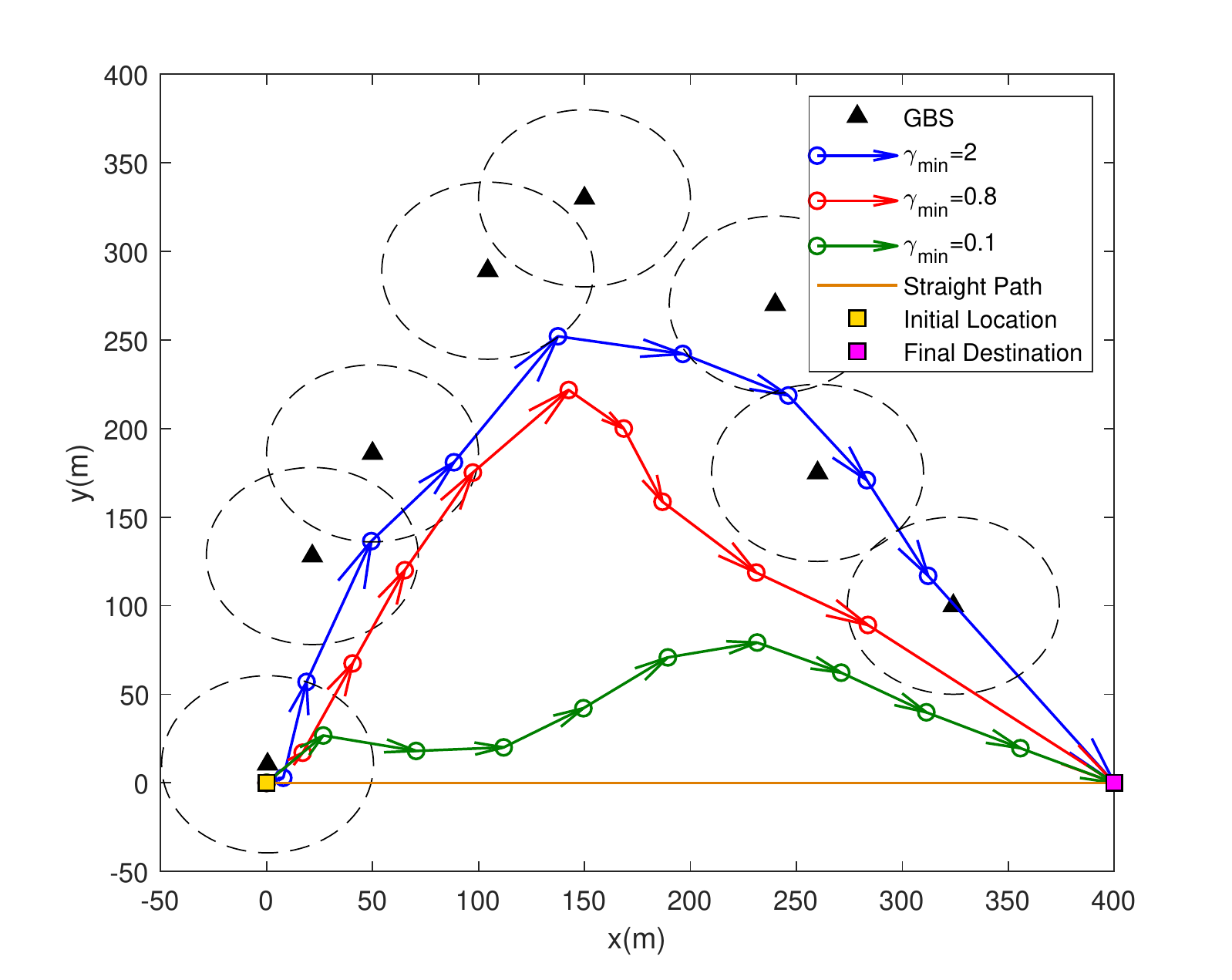}
	\vspace{-0.3em}
	\caption{Trajectory of the aerial vehicle for map 2.}
	\label{traj_2}
	\vspace{-1.5em}
\end{figure}%

Fig. \ref{traj_1} and Fig. \ref{traj_2} show the trajectory of the aerial user for two maps under different values of $\gamma_{\text{min}}$. The initial location of the trajectory of the aerial vehicle in both maps is ${\bf{Q}}^0=(0,0)$. The final position of the aerial vehicle in map 1 and 2 are ${\bf{Q}}^F=(100,400)$ and ${\bf{Q}}^F=(400,0)$, respectively. The dashed circles in these figures show the boundary of regions where the received SNR of the aerial vehicle from the corresponding GBS is no less than $2$. As can be seen, when $\gamma_{\text{min}}=2$, the aerial vehicle must choose its locations inside the circles at time $t_n$ for $\forall n$. In other words, for high $\gamma_{\text{min}}$ regimes, to satisfy the cellular-connectivity constraint, the aerial user needs to fly closer to the GBSs. In contrast, when $\gamma_{\text{min}}$ decreases, the aerial vehicle has more freedom to select its trajectory points from a wider feasibe area. Therefore, the length of its trajectory is reduced. In particular, as $\gamma_{\text{min}}$ becomes closer to $0$, the trajectory of the aerial user tends to be closer to the straight line from the initial location to the final destination. This figure also shows that in general, the aerial vehicle does not have to satisfy the minimum received SINR constraints in times other than $t_n$, $\forall n$. Since the maximum time duration that the aerial vehicle can be disconnected from the cellular service is $T_c$, we can guarantee that the cellular connectivity constraint is satisfied.

\begin{figure}
	\centering
	\includegraphics[width=2.7in,keepaspectratio]{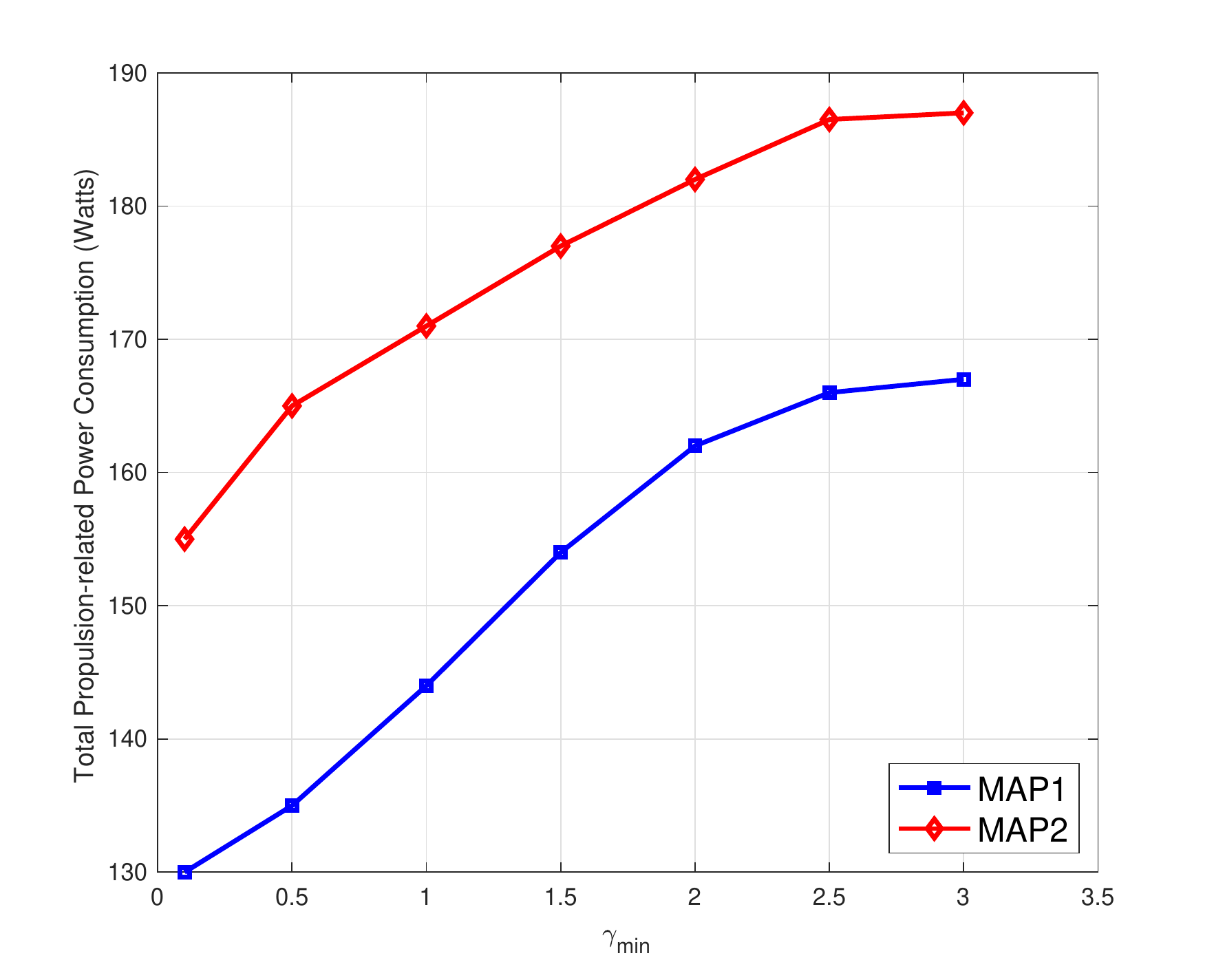}
	\vspace{-0.3em}
	\caption{Total propulsion-related power consumption of the aerial vehicle versus the minimum required SINR $\gamma_{\text{min}}$.}
	\label{fig_p_vs_gamma}
	\vspace{-1.5em}
\end{figure}

Fig. \ref{fig_p_vs_gamma} presents the total propulsion-related power consumption of the aerial user versus the minimum required SINR for the cellular-connectivity. As discussed, when $\gamma_{\text{min}}$ is small, the trajectory of the aerial user is closer to straight line between the initial point and the final destination. Therefore, the length of the trajectory is small and its propulsion power consumption is consequently low. The length of the trajectory and hence, the propulsion power consuption grow as the minimum required SINR increases. As $\gamma_{\text{min}}$ becomes significantly large, it is highly possible that the constraints of \eqref{Problem_original}, including cellular-connectivity one, are not satisfied and hence, the problem does not have any feasible solution. As a result, the increasing trend of the power consumption curve will stop afterwards.  

\section{Conclusion}
In this paper, we addressed the trajectory optimization problem for an aerial vehicle. The objective of the problem is to minimize the total propulsion-related power consumption of the aerial vehicle, ensuring that the cellular-connectivity constraint is satisfied. This problem is a non-convex mixed integer non-linear problem. To obtain an efficient solution for this challenging optimization problem, first, the problem is relaxed and reformulated to a more tractable mathematical form. Then, based on the successive convex approximation technique, an iterative algorithm is developed to convert the problem into a sequence of convex problems. Simulation results show that the proposed algorithm works well and converges in a few number of iterations.

\vspace{-0.4em}
\bibliographystyle{IEEEtran}
\bibliography{Citations}

\begin{thebibliography}{10}
\providecommand{\url}[1]{#1}
\csname url@samestyle\endcsname
\providecommand{\newblock}{\relax}
\providecommand{\bibinfo}[2]{#2}
\providecommand{\BIBentrySTDinterwordspacing}{\spaceskip=0pt\relax}
\providecommand{\BIBentryALTinterwordstretchfactor}{4}
\providecommand{\BIBentryALTinterwordspacing}{\spaceskip=\fontdimen2\font plus
\BIBentryALTinterwordstretchfactor\fontdimen3\font minus
  \fontdimen4\font\relax}
\providecommand{\BIBforeignlanguage}[2]{{%
\expandafter\ifx\csname l@#1\endcsname\relax
\typeout{** WARNING: IEEEtran.bst: No hyphenation pattern has been}%
\typeout{** loaded for the language `#1'. Using the pattern for}%
\typeout{** the default language instead.}%
\else
\language=\csname l@#1\endcsname
\fi
#2}}
\providecommand{\BIBdecl}{\relax}
\BIBdecl

\bibitem{R_Zhang_Survey}
Y.~{Zeng}, R.~{Zhang}, and T.~J. {Lim}, ``{Wireless Communications with
  Unmanned Aerial Vehicles: Opportunities and Challenges},'' \emph{IEEE Commun.
  Mag.}, vol.~54, no.~5, pp. 36--42, May 2016.

\bibitem{Perspective_RZhang}
S.~{Zhang}, Y.~{Zeng}, and R.~{Zhang}, ``{Cellular-Enabled UAV Communication: A
  Connectivity-Constrained Trajectory Optimization Perspective},'' \emph{IEEE
  Trans. Commun.}, vol.~67, no.~3, pp. 2580--2604, Mar. 2019.

\bibitem{Survey_Implementation}
S.~{Chandrasekharan}, K.~{Gomez}, A.~{Al-Hourani}, S.~{Kandeepan},
  T.~{Rasheed}, L.~{Goratti}, L.~{Reynaud}, D.~{Grace}, I.~{Bucaille},
  T.~{Wirth}, and S.~{Allsopp}, ``{Designing and implementing future aerial
  communication networks},'' \emph{IEEE Commun. Mag.}, vol.~54, no.~5, pp.
  26--34, May 2016.

\bibitem{Energy_tradeoff}
D.~{Yang}, Q.~{Wu}, Y.~{Zeng}, and R.~{Zhang}, ``{Energy Tradeoff in
  Ground-to-UAV Communication via Trajectory Design},'' \emph{IEEE Trans.
  Vehicular Tech.}, vol.~67, no.~7, pp. 6721--6726, Jul. 2018.

\bibitem{MEC_RZhang}
\BIBentryALTinterwordspacing
X.~Cao, J.~Xu, and R.~Zhang, ``{Mobile Edge Computing for Cellular-Connected
  {UAV:} Computation Offloading and Trajectory Optimization},'' 2018. [Online].
  Available: \url{http://arxiv.org/abs/1803.03733}
\BIBentrySTDinterwordspacing

\bibitem{Guevenc}
E.~{Bulut} and I.~{Guevenc}, ``{Trajectory Optimization for Cellular-Connected
  UAVs with Disconnectivity Constraint},'' in \emph{IEEE ICC Workshops}, May
  2018, pp. 1--6.

\bibitem{Noma}
W.~{Mei} and R.~{Zhang}, ``{Uplink Cooperative NOMA for Cellular-Connected
  UAV},'' \emph{IEEE J. Sel. Topics Signal Process}, pp. 1--1, 2019.

\bibitem{Saad}
U.~{Challita}, W.~{Saad}, and C.~{Bettstetter}, ``{Deep Reinforcement Learning
  for Interference-Aware Path Planning of Cellular-Connected UAVs},'' in
  \emph{IEEE ICC}, May 2018, pp. 1--7.

\bibitem{TWC_joint_R_Zhang}
Q.~{Wu}, Y.~{Zeng}, and R.~{Zhang}, ``{Joint Trajectory and Communication
  Design for Multi-UAV Enabled Wireless Networks},'' \emph{IEEE Trans. Wireless
  Commun.}, vol.~17, no.~3, pp. 2109--2121, Mar. 2018.

\bibitem{TWC_Energy_Efficient_R_Zhang}
Y.~{Zeng} and R.~{Zhang}, ``{Energy-Efficient UAV Communication With Trajectory
  Optimization},'' \emph{IEEE Trans. Wireless Commun.}, vol.~16, no.~6, pp.
  3747--3760, Jun. 2017.

\bibitem{Behzad_CommL}
B.~{Khamidehi}, A.~{Rahmati}, and M.~{Sabbaghian}, ``{Joint Sub-Channel
  Assignment and Power Allocation in Heterogeneous Networks: An Efficient
  Optimization Method},'' \emph{IEEE Commun. Lett.}, vol.~20, no.~12, pp.
  2490--2493, Dec. 2016.

\bibitem{convexOpt}
S.~Boyd and L.~Vandenberghe, \emph{Convex Optimization}.\hskip 1em plus 0.5em
  minus 0.4em\relax {Cambridge University Press}, 2004.

\end{thebibliography}

\end{document}